\documentclass[a4paper,11pt]{article}
\pdfoutput=1 

\usepackage{jcappub} 

\usepackage[T1]{fontenc} 

\title{\boldmath Constraints on kinematic parameters at $z\ne0$}

\author[a,b]{C. Rodrigues Filho}
\author[b]{Ed\'esio M. Barboza Jr.,\note{Corresponding author.}}

\affiliation[a]{Departamento de F\'{i}sica Te\'orica e Experimental, Universidade Federal do Rio Grande do Norte,\\ 59072-970, Natal-RN, Brazil}
\affiliation[b]{Departamento de F\'isica, Universidade do Estado do Rio Grande do Norte,\\ 59610-210, Mossor\'o - RN, Brazil}

\emailAdd{coenelio@fisica.ufrn.br}
\emailAdd{edesiobarboza@uern.br}

\abstract{The standard cosmographic approach consists in performing a series expansion of a cosmological observable around $z=0$ and then using the data to constrain the cosmographic (or kinematic) parameters at present time. Such a procedure works well if applied to redshift ranges inside the $z$-series convergence radius ($z<1$), but can be problematic if we want to cover redshift intervals that fall outside the $z-$series convergence radius. This problem can be circumvented if we work with  the $y-$redshift, $y=z/(1+z)$, or the scale factor, $a=1/(1+z)=1-y$, for example. In this paper, we use the scale factor $a$ as the variable of expansion. We expand the luminosity distance and the Hubble parameter around an arbitrary $\tilde{a}$ and use the Supernovae Ia (SNe Ia) and the Hubble parameter data to estimate $H$, $q$, $j$ and $s$ at $z\ne0$ ($\tilde{a}\neq1$). We show that the last relevant term for both expansions is the third. Since the third order expansion of $d_L(z)$ has one parameter less than the third order expansion of $H(z)$, we also consider, for completeness, a fourth order expansion of $d_L(z)$. For the third order expansions, the results obtained from both SNe Ia and $H(z)$ data are incompatible with the $\Lambda$CDM model at $2\sigma$ confidence level, but also incompatible with each other. When the fourth order expansion of $d_L(z)$ is taken into account, the results obtained from SNe Ia data are compatible with the $\Lambda$CDM model 
at $2\sigma$ confidence level, but still remains incompatible with results obtained from $H(z)$ data. 
These conflicting results may indicate a tension between the current SNe Ia and $H(z)$ data sets.}

\keywords{cosmography, cosmic chronometers, SN Ia luminosity distances}

\begin{document}
\maketitle
\flushbottom

\section{Introduction}
\label{intro}

By the end of 20th century it was discovered that the Universe is expanding at an accelerating rate 
\cite{Riess,Perlmutter}. 
The current cosmic acceleration can be explained by the existence of a positive cosmological 
constant in the Einstein field equations \cite{LCDM}. However, the cosmological constant presents a huge 
discrepancy between its observed and its theoretical value \cite{CC_problem}. 
Modifications of gravity 
theory \cite{brane1,brane2,brane3,brane4,fr1,fr2,fr3} and exotic forms of fields \cite{quint,phant,quintom} 
are some alternatives to the cosmological constant to 
explain the cosmic acceleration. However, the information about the cosmological parameters 
obtained from these alternative scenarios largely depends on the model under consideration. 

Cosmokinetics (or cosmography) \cite{zcosmog1,zcosmog2,zcosmog3,cosmography1,cosmography2,cosmography3} 
is the least model-dependent method to get information about 
Universe expansion history. The basic assumption of cosmokinetics is the 
cosmological principle.  
No assumptions about sources or gravity theory are made. Therefore, it 
is expected that the results obtained from this kinematic approach remain 
valid regardless of the underlying cosmology. This feature may be an efficient weapon 
to probe the viability of several cosmological models proposed to describe the current 
phase of accelerated expansion of the Universe. For instance, since $j(z)=1$ for the $\Lambda$CDM model,
we can rule out this model if we find that $j\neq1$. 

Cosmography methodology 
consists of expanding cosmological observables such as the Hubble parameter and the 
luminosity distance in power series. However, to obtain some information about 
the kinematic state of the Universe, these series should be stopped. In such an 
approximate process issues arise concerning the series convergence and the 
series truncation order. The series convergence problem can be circumvented by 
choosing a suitable expansion variable such as the so-called $y-$redshift, 
$y=z/(1+z)$ \cite{Cattoen2007a,Cattoen2007b}, or the scale factor, 
$a=(1+z)^{-1}=1-y$ \cite{Barboza2012}, instead of the $z$ redshift. The series 
truncation problem can be alleviated by performing the so-called $F$-test 
\cite{Cattoen2007a,Cattoen2007b,Vitagliano} to find which truncation order 
provides the more statistically significant fit to a given data set.

In this paper we follow the procedure adopted in \cite{Barboza2012} and 
perform the series expansion of the luminosity distance, $d_L(a)$, and of the Hubble 
parameter $H(a)$ around an arbitrary scale factor $\tilde{a}$. The $F-$ test
indicates that the most statistically significant truncation order for both series 
is the third. Since the third order approximation of $d_L$ has one parameter less than 
the third order truncation of $H$, we also consider, for sake of completeness, the fourth order $d_L$ approximation. 
We use some of the most 
recent Type Ia Supernovae (SNe Ia) and $H(z)$ data sets to 
constrain the Hubble ($H$), deceleration ($q$), jerk ($j$) and snap ($s$) 
kinematic parameters at $z\ne0$. 
For the third order expansions, the results obtained from both SNe Ia and $H(z)$ data are incompatible with the $\Lambda$CDM model at $2\sigma$ confidence level, but also incompatible with each other. When the fourth order expansion of $d_L$ is taken into account, the results obtained from SNe Ia data are compatible with the $\Lambda$CDM model 
at $2\sigma$ confidence level, but still remains incompatible with results obtained from $H(z)$ data.
The constraints on $j$ and $s$ are conflicting and indicate a 
discrepancy between SNe Ia and $H(z)$ measurements.

\section{Cosmokinetics}

Cosmokinetics relies on the assumption that at large scales the Universe is 
homogeneous and isotropic. Mathematically, this assumption is translated by the 
Robertson-Walker (RW) metric

\begin{equation}
\label{RW}
ds^2=-c^2dt^2+a^2(t)\Big[\frac{dr^2}{1-kr^2}+r^2(d\theta^2+\sin^2\theta\, 
d\phi^2)\Big],
\end{equation}

\noindent where $a(t)$ is the scale factor of the Universe and $k$ is the 
Universe spatial curvature. In agreement with recent results of the CMB power 
spectrum~\cite{Planck}, we restrict our attention to a spatially flat Universe 
($k = 0$) in this paper. For a flat RW line element, the luminosity distance 
takes the form,

\begin{equation}
\label{Dl-flatRW}
d_L=c\,\frac{1}{a}\,\int_{t}^{t_0}\frac{dt^{\prime}}{a(t^{\prime})}=c\,(1+z)
\int_{0}^{z}\frac{dz^{\prime}}{H(z^{\prime})},
\end{equation}

\noindent where the subscript $0$ denotes the value of a variable at the present
epoch, $H\equiv a^{-1}(da/dt)$ is the Hubble parameter, which provides the expansion 
rate of the Universe, and we have used the convention $a_0=1$. 

Cosmokinetics works at time domains where a complete knowledge of the $a(t)$ 
function is not necessary. The standard approach consists in performing a Taylor 
expansion of a cosmological observable in terms of the redshift, keeping the 
expansion center fixed at $z=0$ \cite{zcosmog1,zcosmog2,zcosmog3}. By focusing 
on the Hubble parameter and the luminosity distance, such a procedure leads to

\begin{eqnarray}
\label{H-expansion}
H(z)&=&H_0\Big[1+(1+q_0)z+\frac{1}{2}(j_0-q_0^2)z^2
       +\frac{1}{6}(3q_0^3+3q_0^2-3j_0+4q_0j_0-s_0)z^3+\nonumber\\
       &+&\frac{1}{24}(l_0+8s_0+7q_0s_0+12j_0+32q_0j_0+25q_0^2j_0
       -4j_0^2-12q_0^2-24q_0^3-\nonumber\\&-&15q_0^4)z^4+\cdots\Big]
\end{eqnarray}

\noindent and

\begin{eqnarray}
\label{Dl-expansion}
d_L(z)&=&\frac{c}{H_0}\Big\{z+\frac{1}{2}(1-q_0)\,z^2
+\frac{1}{6}(3q_0^2+q_0-1-j_0)z^3+
\frac{1}{24}(2-2q_0-15q_0^2-\nonumber\\
&-&15q_0^3+5j_0+10q_0j_0+s_0)z^4-
\frac{1}{120}[6(1-q_0)-3q_0^2(27+55q_0-35q_0^2)+\nonumber\\
&+&5q_0j_0(21q_0+22)+15q_0s_0-10j_0^2+27j_0+11s_0+l_0]z^5+\cdots\Big\},
\end{eqnarray}
where
\begin{equation}
\label{parameters}
q\equiv-\frac{\ddot{a}}{H^2a},\ \ j\equiv\frac{\dddot{a}}{H^3a},
\ \ s\equiv\frac{\ddddot{a}}{H^4a} \ \ \mbox{and}\ \ l\equiv\frac{a^{(5)}}{H^5a}
\end{equation}
are, respectively, the deceleration, the jerk, the snap and the lerk parameters, and the 
dot denotes time derivatives. These parameters provide information about the 
kinematic state of the Universe. Physically, $q$ specifies if the Universe is 
expanding at an accelerated ($q<0$), decelerated  ($q>0$) or constant ($q=0$) 
rate; $j$ shows whether the Universe's acceleration is increasing ($j>0$), decreasing 
($j<0$) or constant ($j=0$); $s$ tells us if $d^3a/dt^3$ is increasing ($s>0$), 
decreasing ($s<0$) or constant ($s=0$) and $l$ tell us if $d^4a/dt^4$ is increasing ($l>0$), 
decreasing ($l<0$) or constant ($l=0$).  Thus, the kinematic approach allow us 
to investigate the cosmic acceleration without assuming modifications of the gravity 
theory or dark energy models. 

The truncation of the expansions 
(\ref{H-expansion}) and (\ref{Dl-expansion}) at the first two or three terms 
should be good approximations if $z$ does not lie outside the convergence 
radius of these series, $z<1$ \cite{Cattoen2007a,Cattoen2007b}. However, 
currently we have measurements of $H$ and $d_L$ at $z>1$. Applying low order 
approximations to cover such a redshift range may result in artificially 
strong constraints on the free parameters, while taking higher order terms, and consequently 
increasing the number of free parameters, can make the analysis more laborious than necessary.
Therefore, we need to find a way to 
cover the higher redshift range using the lowest number of parameters possible. 
This problem can be handled if we work with the $y$-redshift, $y=z/(1+z)$  
\cite{Cattoen2007a,Cattoen2007b} which maps the redshift domain 
$z\in[0,\infty[$ into $y\in[0,1[$\, or, equivalently, if we work with the scale 
factor $a$ ($a=1-y$) \cite{Barboza2012} as expansion variables. Here we choose the scale factor as the  
expansion variable. Note that an expansion around $z=0$ is translated to an 
expansion around $a=1$ when the scale factor is used as the expansion variable. The 
standard approach consists in taking the expansion center at $z=0$ ($a=1$). However, 
nothing prevents us from changing the expansion center to an arbitrary redshift or 
scale factor. By assuming that the Hubble parameter and the luminosity distance 
are analytical functions in the range $]\tilde{a}-\epsilon,\tilde{a}+\epsilon[$, 
where $\tilde{a}$ is expansion center, we get

\begin{table}[tpb]
\centering
\begin{tabular}{lccccccl}
\hline
\hline
Order&$H_0$ & $q_0$ & $j_0$ &$s_0$ &$l_0$&$\chi^2_{min}$& Data\\
\hline
2nd&$73.9^{+4.8}_{-4.8}$&$-1.73^{+0.70}_{-0.64}$&$13.1^{+6.4}_{-5.8}$&-&-&15.53&$H(z)$\\
3rd&$73.8^{+4.7}_{-4.7}$&$-0.92^{+1.49}_{-1.41}$&$-0.4^{+21.6}_{-18.4}$&$-45.0^{+263}_{-86.0}$&-&13.44&\\
4th&$72.9^{+4.7}_{-4.7}$&$0.16^{+3.04}_{-3.24}$ &$-23.4^{+77}_{-57}$&$-256.0^{+1256}_{-194}$&$1080^{+38020}_{-3580}$&13.06&\\
\hline
3rd&$69.98_{-1.08}^{+1.08}$&$-0.51_{-0.39}^{+0.38}$&$-0.60_{-3.9}^{+4.8}$&-&-&$562.19$&$\mu(z)$\\
4th&$69.97_{-1.41}^{+1.43}$&$-0.50_{-0.90}^{+0.86}$&$-0.71_{-14.43}^{+19.24}$&$-18.8_{-83.59}^{+321.2}$&-&$562.19$&\\
\hline
\hline
\end{tabular}
\caption{\label{chi_square_cosmography} Constraints on $H,\,q,\,j,\,{\rm and}\,s$ at $z=0$ obtained from $H$ and $d_L$ 
measurements for successive approximation orders of the expansions (\ref{H-a-general}) and (\ref{Dl-a-general}). 
The errors correspond to $2\sigma\,(\Delta\chi^2=4)$ statistical uncertainty for a single parameter. 
The $\chi^2_{min}$ values shows that the relevant approximation order for both expansions is the 3rd.}
\end{table}

\begin{eqnarray}
\label{H-a-general}
H(a)&=&\tilde{H}\Big\{1+(1+\tilde{q})\Big(1-\frac{a}{\tilde{a}}\Big)+\frac{1}{2}(2+2\tilde{q}-\tilde{q}^2+\tilde{j})\Big(1-\frac{a}{\tilde{a}}
\Big)^2-\frac{1}{6}(\tilde{s}-3\tilde{j}+3\tilde{q}^2-3\tilde{q}^3-\nonumber\\
&-&6\tilde{q}+4\tilde{q}\tilde{j}-6)\Big(1-\frac{a}{\tilde{a}}\Big)^3+
\frac{1}{24}[\tilde{l}+7\tilde{q}\tilde{s}+5\tilde{q}^2(5\tilde{j}-3\tilde{q}^2)-4\tilde{j}^2-
4(\tilde{s}-3\tilde{j}+3\tilde{q}^2-\nonumber\\&-&3\tilde{q}^3-6\tilde{q}+4\tilde
{q}\tilde{j}-6)]\Big(1-\frac{a}{\tilde{a}}\Big)^4+\cdots \Big\}
\end{eqnarray}

\noindent and

\begin{eqnarray}
\label{Dl-a-general}
\frac{\tilde{H}d_L(a)}{c}&=&\frac{1}{a}\Big\{\tilde{a}
\frac{\tilde{H}\tilde{d}_L}{c}+\frac{1}{\tilde{a}}\Big(1-\frac{a}{\tilde{a}}
\Big)\Big[1 +
\frac{1}{2}(1-\tilde{q})
\Big(1-\frac{a}{\tilde{a}}\Big)+
\frac{1}{6}(2-2\tilde{q}+3\tilde{q}^2-\tilde{j})\Big(1-\frac{
a}{\tilde{a}}\Big)^2+\nonumber\\
&+&\frac{1}{24}(\tilde{s}-3\tilde{j}+9\tilde{q}^2-15\tilde{q}
^3-6\tilde{q}+10\tilde{q}\tilde{j}+6)
\Big(1-\frac{a}{\tilde{a}}
\Big)^3-\frac{1}{120}(\tilde{l}+15\tilde{q}\tilde{s}+105\tilde{q}^2\tilde{j}-\nonumber\\
&-&10\tilde{j}^2-
105\tilde{q}^4-4\tilde{s}+12\tilde{j}-36\tilde{q}^2+60\tilde{q}
^3+24\tilde{q}-
40\tilde{q}\tilde{j}-24)\Big(1-\frac{a}{\tilde{a}}\Big)^4\Big]\cdots \Big\},
\end{eqnarray}

\noindent where a tilde denotes a function evaluated at $\tilde{a}$. The main 
advantage of this procedure is that  we can estimate the value of the 
cosmographic  parameters at $z\ne0$ and so, changing the expansion center, 
discover how these parameters evolve in a completely model-independent way. Note 
that  $\tilde{d}_L=d_L(\tilde{a})$ is also a free parameter in our cosmographic 
analysis. Since $d_L=0$ at $a=1$, we can write $\tilde{d}_L$ in terms of 
$\tilde{H}$, $\tilde{q}$, $\tilde{j}$ and so on. 
Thus, by expanding $H$ and $d_L$ around an arbitrary scale factor $\tilde{a}$ it is 
possible to obtain the cosmographic parameters as a function of $a$ independent of the 
underlying cosmological model. Also, it is worth mentioning that the lower the 
value of $\epsilon$ the better the approximation that describes the real $H$ and $d_L$ 
functions.

\section{Observational constraints}

\subsection{Data}

In order to constrain the cosmographic parameters we use separately the $580$ SNe Ia 
distance measurements of the Union 2.1 compilation \cite{Union2.1} and the $30$ 
measurements of the Hubble parameter compiled in \cite{Moresco2016}, plus the 
measurement of the Hubble constant $H_0=73.24\pm1.74$ ${\rm Km}\cdot{\rm s}^{-1}\cdot{\rm Mpc}^{-1}$ 
provided by \cite{RiessH_0}. The SNe Ia data are distributed in the redshift interval 
$0.015\leq z \leq 1.414$ ($0.414\leq a\leq 0.985$), corresponding to a maximum 
$\epsilon$ of $\sim 0.571$, while the Hubble parameter data cover the redshift range
$0\leq z \leq 1.965$ ($0.337\leq a\leq 1$), corresponding to a maximum 
$\epsilon$ of $\sim 0.663$. 

For SNe Ia data, the statistical analysis is performed using the distance modulus definition:
\begin{eqnarray}
\mu(z\vert \{\theta_i\})&=&5\log_{10} d_L(z\vert \{\theta_i\})+25\nonumber\\
&=&5\log_{10} \Big(\frac{\tilde{H}d_L}{c}\Big)-5\log_{10}\Big(\frac{\tilde{h}}{3}\Big)+40,
\end{eqnarray}
where $\{\theta_i\}=\{\tilde{H},\,\tilde{q},\,\tilde{j},\,\dots\}$ 
is the set of parameters to be fitted and 
$\tilde{h}=\tilde{H}/(100 {\rm Km}\cdot{\rm s}^{-1}\cdot{\rm Mpc}^{-1})$. 
The best fit parameters are obtained by minimizing the quantity
\begin{equation}
\chi^2_{{\rm SN}}(\{\theta_i\})=\sum_{i=1}^{580}\frac{[\mu(z\vert \{\theta_i\})-
\mu^{{\rm obs}}(z_i)]^2}{\sigma_{\mu,i}^2},
\end{equation}
where $\mu^{{\rm obs}}(z_i)$ is the observed value of the distance moduli at 
redshift $z_i$ and $\sigma_{\mu,i}^2$ is the error of $\mu^{{\rm obs}}(z_i)$.

For the Hubble parameter data, the best fit parameters are obtained by minimizing 
the quantity
\begin{equation}
\chi^2_H(\{\theta_i\})=\sum_{i=1}^{31}\frac{[H(z\vert \{\theta_i\})-
H^{{\rm obs}}(z_i)]^2}{\sigma_{H,i}^2},
\end{equation}
where $H^{{\rm obs}}(z_i)$ is the observed value of the Hubble parameter at 
$z_i$ and $\sigma_{H,i}^2$ is the error associated with the $H^{{\rm obs}}(z_i)$ 
measurement.

\subsection{F-test}

In order to decide the order in which the series should be stopped, 
we perform the so-called $F-$test, defined as
\begin{equation}
F_{kl}=\frac{\chi_k^2-\chi_l^2}{n_l-n_k}\frac{\chi_l^2}{N-n_l},
\end{equation}
where $\chi_i^2$ and $n_i$ are, respectively, the minimum chi-squared 
function and the number of parameters of the $i$th model and $N$ is 
the number of data points. This test compares two models, identifying the 
one that provides the best fit to the data, with the null hypothesis implying the 
correctness of the first model. In the following we compare successive 
truncations of the Taylor series (\ref{H-a-general}) and (\ref{Dl-a-general}) 
to decide the number of parameters that we need to take into account in our analysis. 
Table \ref{chi_square_cosmography} displays the constraints on the cosmographic 
parameters at the present time for successive approximations of $H$ and $d_L$. It is 
easy to see that for both expansions the last relevant term is the third, 
$F_{34}\approx0.2$ for $H$ and $F_{34}=0$ for $d_L$. However, the third order 
approximation of $H$ contains four parameters while the third order approximation 
of $d_L$ contains three parameters. Therefore, for sake of completeness, 
we will also work with one term beyond than necessary in 
the $d_L$ series approximation.
In what follows we take the third order
approximation of $H$ and the third and fourth order
approximation of $d_L$ and compare the constraints on 
$H,\,q,\,j,\,{\rm and}\,s$ obtained from $H(z)$ and SNe Ia data.

\subsection{Results}

The evolution of the cosmographic parameters $H,\,q,\,j\,{\rm and}\, s$ is obtained 
following the algorithm:
\begin{enumerate}
\item[1.] fix the expansion center $\tilde{a}_i=(1+\tilde{z}_i)^{-1}$ in eqs. 
 (\ref{H-a-general}) and (\ref{Dl-a-general});
\item[2.] perform the statistical analysis with $H$ and SNe Ia data to constrain 
 $H,\,q,\,j\,{\rm and}\, s$ at $\tilde{z}_i$;
\item[3.] set $\tilde{z}_{i+1}=\tilde{z}_i+\Delta\tilde{z}$ and repeat the previous step 
 to constrain $H,\,q,\,j\,{\rm and}\, s$ at $\tilde{z}_{i+1}$.
\end{enumerate}
Here we take a step of $\Delta\tilde{z}=0.1$ and cover the interval $0\leq\tilde{z}_i\leq1.4$ 
for both data sets used in our analysis.

Table \ref{parameters_H} contain the results obtained from $H$ data.
Tables \ref{parameters_SN1} and \ref{parameters_SN} contain 
the results obtained from SNe Ia data for the third and fourth order approximations, respectively. The errors correspond to a  
$2\sigma\, (\Delta\chi^2=4)$ confidence interval for each parameter. 
In all cases the reduced chi-square values ($\chi^2_{\nu}=\chi^2_{min}/{\rm NDoF}$) 
remain unchanged when the expansion center is shifted.

\begin{table}[tbp]
\centering
 \begin{tabular}{ccccc}
 \hline
 \hline
 $z$ & $H$ & $q$ & $j$ & $s$\\ 
 \hline
    0.0 & $73.8_{-4.7}^{+4.7}$&  $-0.92_{-1.41}^{+1.49}$&  $-0.40_{-18.4}^{+21.6}$& $-45.0_{-86.0}^{+263}$\\
    0.1 & $74.3_{-5.0}^{+5.0}$&  $-0.87_{-0.47}^{+0.43}$&  $2.50_{-8.70}^{+9.80 }$&$-15.0_{-85.0}^{+110}$ \\
    0.2 & $75.8_{-5.2}^{+5.2}$&  $-0.66_{-0.49}^{+0.42}$&   $3.21_{-3.41}^{+3.99}$& $-2.99_{-47.0}^{+45.0}$ \\
    0.3 & $78.8_{-4.9}^{+5.0}$&  $-0.39_{-0.48}^{+0.47}$&   $3.16_{-1.66}^{+1.94}$&  $1.50_{-22.5}^{+21.5}$ \\
    0.4 & $83.1_{-5.6}^{+5.6}$&  $-0.16_{-0.46}^{+0.41}$&   $2.76_{-1.80}^{+2.16}$&  $2.80_{-10.0}^{+12.8}$ \\
    0.5 & $88.6_{-6.8}^{+6.5}$&   $0.01_{-0.32}^{+0.32}$&   $2.25_{-2.08}^{+2.43}$&   $3.30_{-5.80}^{+15.5}$\\
    0.6 & $94.5_{-7.3}^{+8.2}$&   $0.14_{-0.26}^{+0.24}$&   $1.70_{-2.10}^{+2.46}$&   $3.90_{-5.10}^{+5.9}$ \\
    0.7 &$102.0_{-8.4}^{+8.4}$&   $0.21_{-0.23}^{+0.24}$&   $1.25_{-2.05}^{+2.35}$&   $4.30_{-4.70}^{+12.4}$\\
    0.8 &$109.6_{-8.9}^{+8.7}$&   $0.25_{-0.27}^{+0.28}$&   $0.80_{-1.85}^{+2.20}$&   $4.70_{-4.00}^{+2.30}$\\
    0.9 &$117.1_{-9.0}^{+9.1}$&   $0.27_{-0.32}^{+0.32}$&   $0.50_{-1.56}^{+1.92}$&   $4.90_{-2.85}^{+1.75}$\\
    1.0 &$125.0_{-9.4}^{+9.3}$&   $0.26_{-0.35}^{+0.35}$&   $0.17_{-1.41}^{+1.71}$&   $4.98_{-2.39}^{+1.07}$\\
    1.1 &$132.9_{-10.0}^{+10.0}$& $0.25_{-0.37}^{+0.38}$&   $-0.06_{-1.21}^{+1.45}$& $4.95_{-3.11}^{+1.01}$\\
    1.2 &$140.7_{-11.1}^{+11.2}$& $0.22_{-0.38}^{+0.40}$&   $-0.27_{-0.99}^{+1.23}$&$4.76_{-3.60}^{+1.24}$\\
    1.3 &$148.5_{-12.7}^{+12.8}$& $0.20_{-0.44}^{+0.38}$&  $-0.42_{-0.81}^{+1.01}$& $4.59_{-3.97}^{+1.50}$\\
    1.4 &$156.3_{-14.9}^{+14.6}$& $0.16_{-0.43}^{+0.38}$&  $-0.58_{-0.62}^{+0.85}$& $4.25_{-4.08}^{+1.93}$\\
 \hline
 \hline
 \end{tabular}
 \caption{ \label{parameters_H} Estimates of $H,\,q,\,j,\,{\rm and}\,s$ as function of the redshift obtained from $H$ measurements for the 
 3rd order approximation of the Hubble parameter (\ref{H-a-general}). The errors correspond to 
 $2\sigma\,(\Delta\chi^2=4)$ statistical uncertainty for each parameter. $\chi^2_{min}=13.44$ for all 
 redshifts considered.}
\end{table}

\begin{table}[tbp]
\centering
 \begin{tabular}{ccccc}
 \hline
 \hline
 $z$ & $H$ & $q$ & $j$ & $s$ \\
 \hline
    0.0 & $69.97_{-1.10}^{+1.10}$&     $-0.51_{-0.39}^{+0.39}$& $-0.60_{-3.90}^{+4.80}$&   $-$\\
    0.1 & $73.28_{-0.88}^{+0.88}$&     $-0.50_{-0.13}^{+0.13}$&  $0.60_{-2.25}^{+2.25}$&   $-$\\
    0.2 & $76.77_{-1.11}^{+1.11}$&     $-0.42_{-0.09}^{+0.09}$&  $1.05_{-1.13}^{+1.10}$&   $-$\\
    0.3 & $80.68_{-1.14}^{+1.14}$&     $-0.33_{-0.13}^{+0.13}$&  $1.19_{-0.54}^{+0.59}$&   $-$\\
    0.4 & $85.15_{-1.52}^{+1.52}$&     $-0.23_{-0.15}^{+0.15}$&  $1.22_{-0.28}^{+0.34}$&   $-$\\
    0.5 & $90.02_{-2.36}^{+2.36}$&     $-0.14_{-0.16}^{+0.16}$&  $1.21_{-0.15}^{+0.15}$&   $-$\\
    0.6 & $95.52_{-3.20}^{+3.28}$&     $-0.05_{-0.16}^{+0.16}$&  $1.20_{-0.10}^{+0.20}$&   $-$\\
    0.7 & $101.32_{-4.14}^{+4.59}$&    $0.02_{-0.17}^{+0.17}$&  $1.20_{-0.08}^{+0.18}$&   $-$\\
    0.8 & $107.76_{-5.40}^{+5.64}$&     $0.09_{-0.16}^{+0.16}$&  $1.21_{-0.07}^{+0.18}$&   $-$\\
    0.9 & $114.50_{-6.60}^{+6.90}$&   $0.15_{-0.15}^{+0.15}$&  $1.23_{-0.07}^{+0.18}$&   $-$\\
    1.0 & $121.33_{-7.48}^{+8.67}$&   $0.19_{-0.15}^{+0.15}$&  $1.25_{-0.07}^{+0.19}$&   $-$\\
    1.1 & $128.60_{-8.60}^{+10.20}$&   $0.24_{-0.12}^{+0.15}$&  $1.27_{-0.07}^{+0.21}$&   $-$\\
    1.2 & $136.75_{-10.00}^{+11.25}$&   $0.28_{-0.12}^{+0.14}$&  $1.31_{-0.09}^{+0.20}$&   $-$\\
    1.3 & $144.48_{-10.92}^{+13.12}$&   $0.32_{-0.11}^{+0.13}$&  $1.34_{-0.09}^{+0.21}$&   $-$\\
    1.4 & $153.16_{-12.32}^{+14.84}$&   $0.35_{-0.10}^{+0.13}$&  $1.37_{-0.10}^{+0.22}$&   $-$\\
 \hline
 \hline
 \end{tabular}
 \caption{\label{parameters_SN1} Estimates of $H,\,q\,{\rm and}\,j$ as function of the redshift obtained from Union 2.1 SNe Ia sample for the 
 3rd order approximation of $d_L$ (\ref{Dl-a-general}). The errors correspond to a 
 $2\sigma\,(\Delta\chi^2=4)$ statistical uncertainty for each parameter.  
 $\chi^2_{\nu}=\chi^2_{min}/{\rm NDoF}=0.974$ for all redshifts considered.}
 \end{table}

\begin{figure}[t]
\centering
\includegraphics[width=16cm,height=8cm]{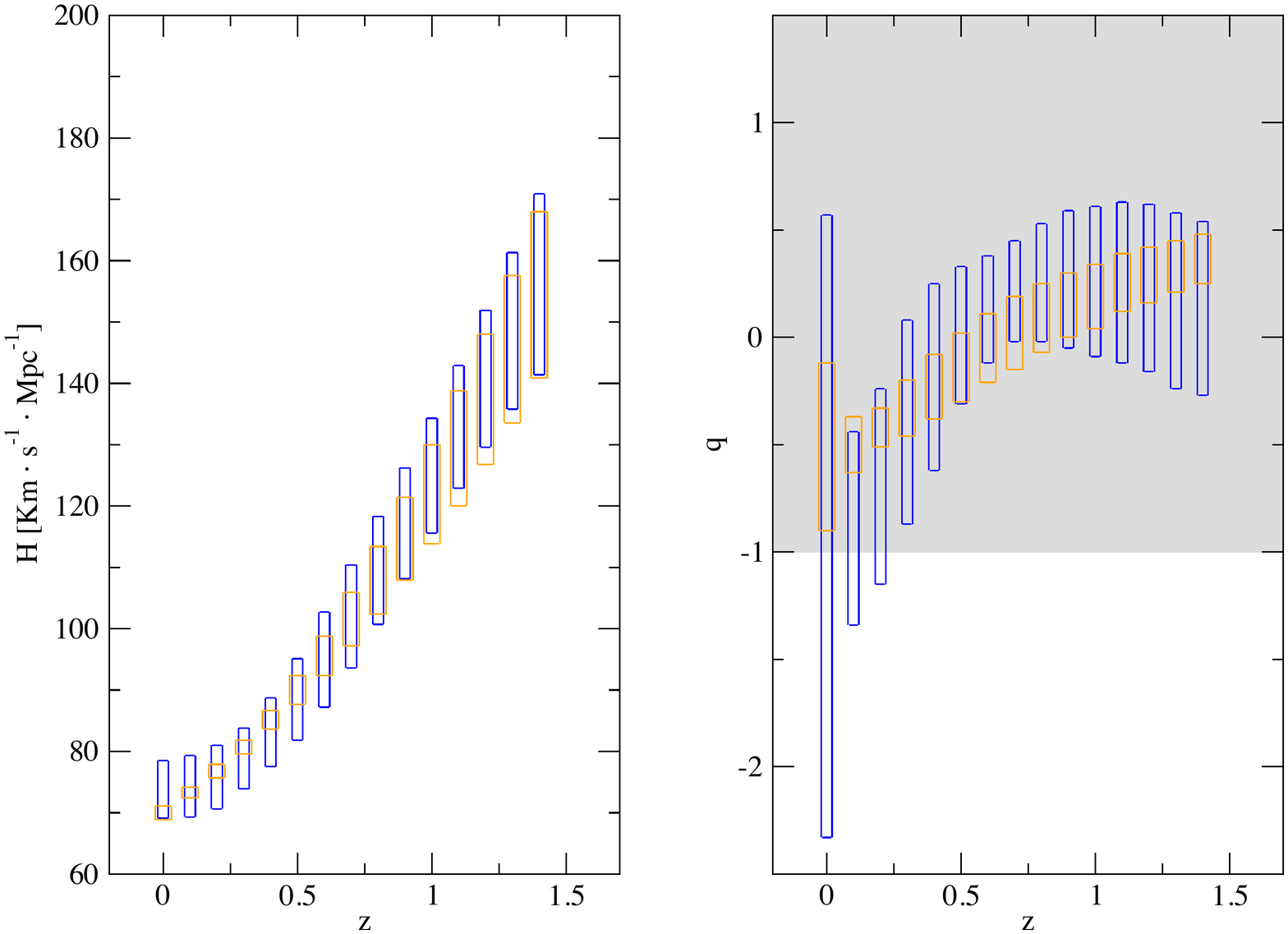}
\caption{\label{3rdH_and_q} Redshift evolution of $H$ (left panel) and $q$ (right panel). The blue boxes corresponds to the constraints 
obtained from $H$ data while the orange boxes corresponds to the constraints obtained from 
SNe Ia data. The gray region represents the region allowed for the $\Lambda$CDM model ($q>-1$).}

\includegraphics[width=16cm,height=8cm]{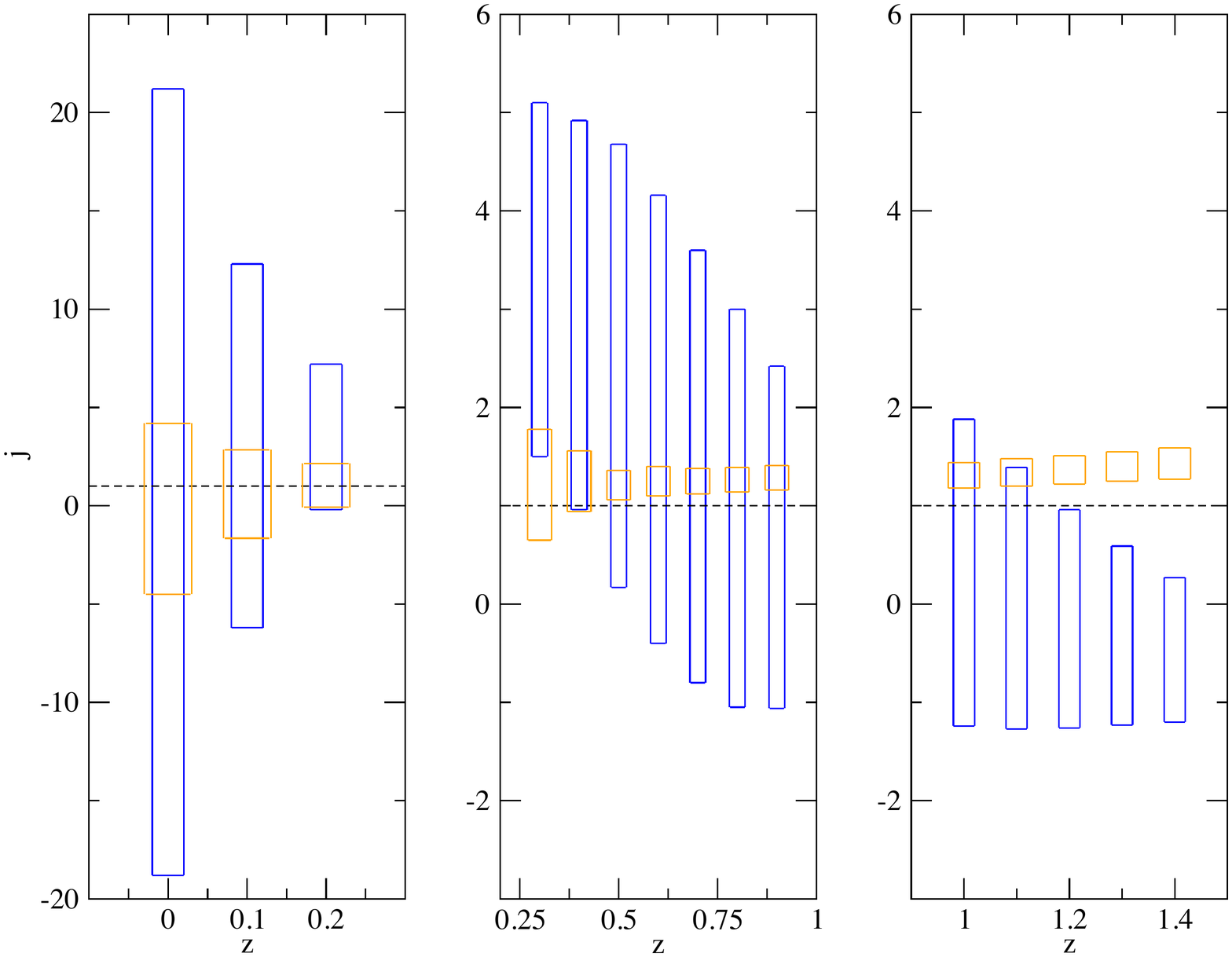}
\caption{\label{3rdj} Redshift evolution of $j$. The blue boxes corresponds to the constraints 
obtained from $H$ data while the orange boxes corresponds to the constraints obtained from 
SNe Ia data. The dashed line represents the $\Lambda$CDM model ($j=1$).}
\end{figure}

\begin{figure}[tbp]
\centering
\includegraphics[width=16cm,height=8cm]{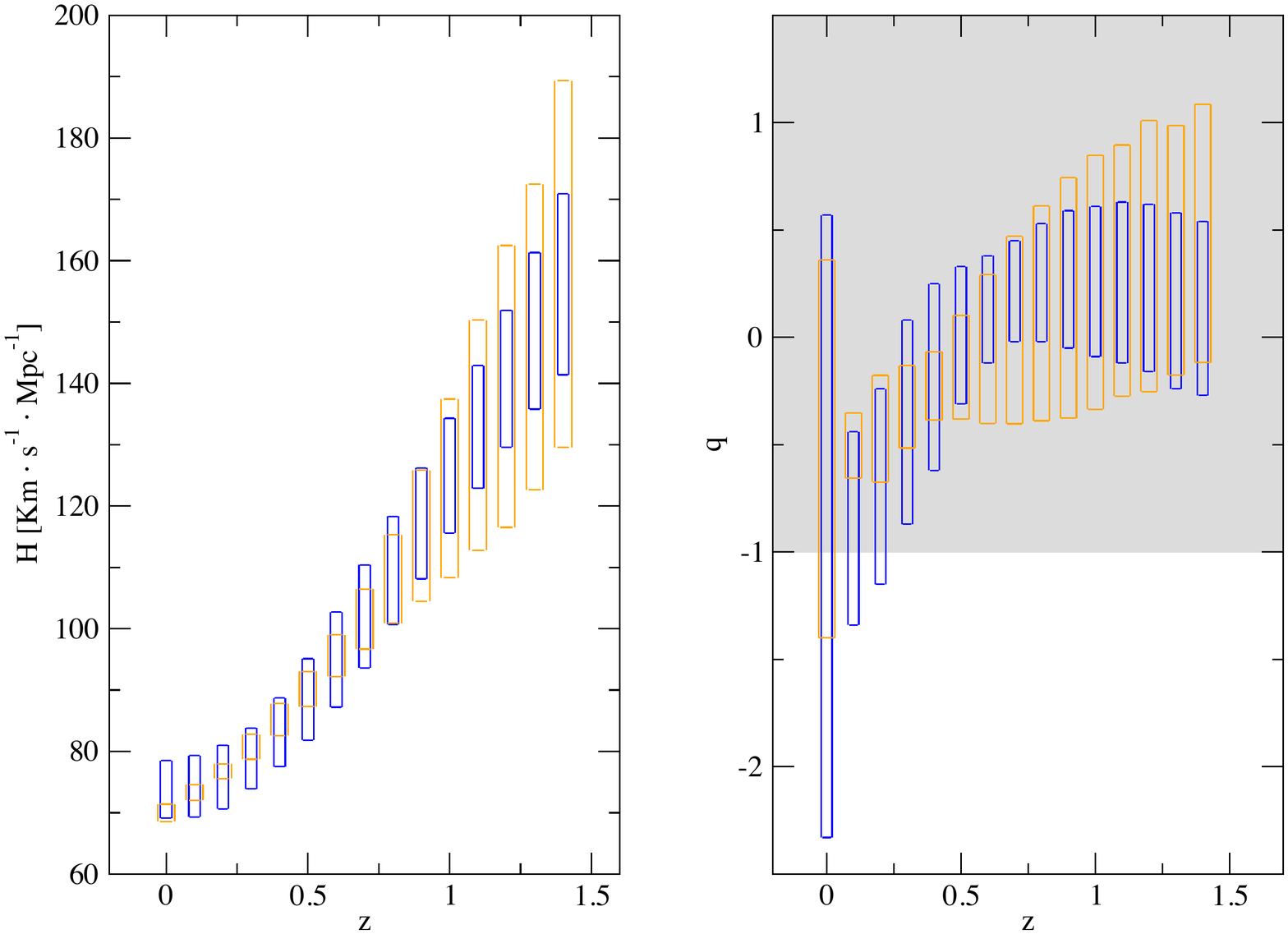}
\caption{\label{H_and_q} Redshift evolution of $H$ (left panel) and $q$ (right panel). The blue boxes corresponds to the constraints 
obtained from $H$ data while the orange boxes corresponds to the constraints obtained from 
SNe Ia data. The gray region represents the region allowed for the $\Lambda$CDM model ($q>-1$).}

\includegraphics[width=16cm,height=8cm]{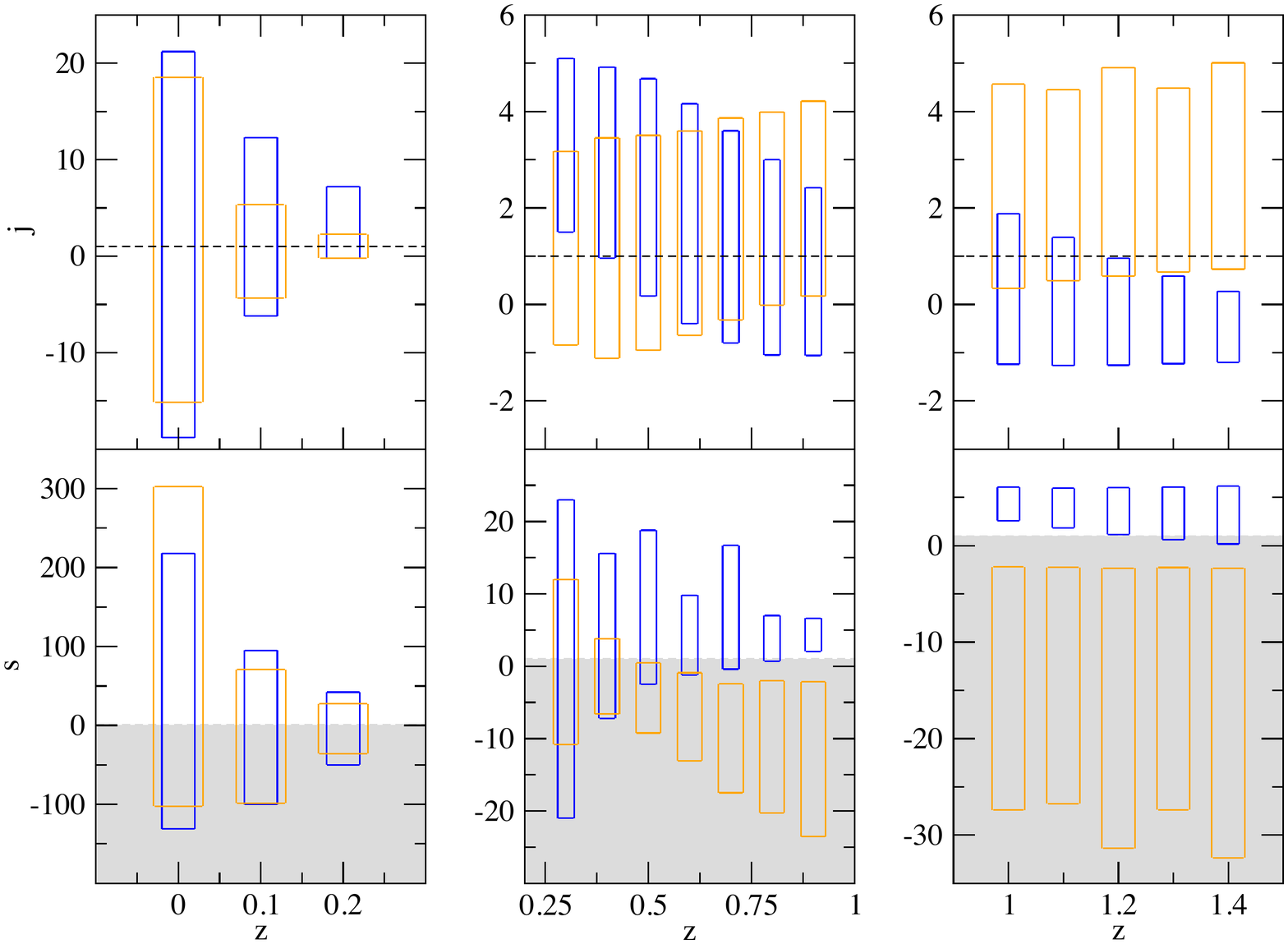}
\caption{\label{j_and_s} Redshift evolution of $j$ (top panels) and $s$ (bottom panels). The blue boxes corresponds to the constraints 
obtained from $H$ data while the orange boxes corresponds to the constraints obtained from 
SNe Ia data. The dashed line represents the $\Lambda$CDM model ($j=1$). The gray region represents the region 
allowed for the $\Lambda$CDM model ($s<1$).}
\end{figure}

In order to make the comparison between these results clearer, a graphical representation of the results contained in Tables \ref{parameters_H} 
and \ref{parameters_SN1} is given in Figs. \ref{3rdH_and_q} and \ref{3rdj} and
a graphical representation of the results contained in Tables \ref{parameters_H} 
and \ref{parameters_SN} is given in Figs. \ref{H_and_q} and \ref{j_and_s}.
Figures \ref{3rdH_and_q} and \ref{3rdj} shows, respectively, the constraints on 
$H$ (left panel of Figure \ref{3rdH_and_q}), $q$ (right panel of Figure \ref{3rdH_and_q}) 
and $j$ (Figure \ref{3rdj}) for the third order approximations of $H$ and $d_L$ at 15 points equally 
spaced in the redshift range $0\leq z\leq1.4$. 
Figure \ref{H_and_q} shows the constraints on $H$ (left panel) and $q$ (right panel) 
and Figure \ref{j_and_s} shows the constraints on $j$ (top panel) and $s$ (bottom panel) 
when a fourth order approximation of $d_L$ is 
considered. 
The blue boxes stand 
for $2\sigma$ confidence intervals obtained from $H$ data while 
the orange boxes stand for $2\sigma$ confidence intervals obtained from 
SNe Ia data. The gray region in the $q$ plots, the dashed line in the $j$ plots and the gray region in the $s$ plots 
correspond, respectively, to the $\Lambda$CDM bounds:
\begin{eqnarray}
 q(a)&=&-1+\frac{3H_0^2\Omega_{m,0}}{2H^2a^3}>-1,\nonumber\\
 j(a)&=&1\quad {\rm and}\\
 s(a)&=&1-\frac{9H_0^2\Omega_{m,0}}{2H^2a^3}<1,\nonumber
\end{eqnarray}
where $\Omega_{m,0}$ is matter density parameter at the present time.

 \begin{table}[t]
\centering
 \begin{tabular}{ccccc}
 \hline
 \hline
 $z$ & $H$ & $q$ & $j$ & $s$ \\
 \hline
    0.0 & $69.97_{-1.41}^{+1.43}$&     $-0.50_{-0.90}^{+0.86}$& $-0.71_{-14.4}^{+19.2}$&   $-18.8_{-83.6}^{+321}$\\
    0.1 & $73.32_{-1.30}^{+1.26}$&     $-0.50_{-0.16}^{+0.14}$&  $0.51_{-4.86}^{+4.83}$&   $-8.40_{-90.0}^{+79.2}$\\
    0.2 & $76.76_{-1.20}^{+1.20}$&     $-0.42_{-0.25}^{+0.25}$&  $1.04_{-1.24}^{+1.24}$&   $-3.25_{-32.3}^{+30.7}$\\
    0.3 & $80.66_{-1.96}^{+2.10}$&     $-0.33_{-0.19}^{+0.20}$&  $1.25_{-2.09}^{+1.93}$&   $-2.40_{-8.40}^{+14.4}$\\
    0.4 & $85.08_{-2.52}^{+2.73}$&     $-0.23_{-0.16}^{+0.16}$&  $1.25_{-2.37}^{+2.20}$&   $-1.70_{-4.90}^{+5.50}$\\
    0.5 & $90.10_{-2.80}^{+2.90}$&     $-0.14_{-0.25}^{+0.24}$&  $1.19_{-2.15}^{+2.31}$&   $-1.75_{-7.48}^{+2.20}$\\
    0.6 & $95.50_{-3.30}^{+3.50}$&     $-0.04_{-0.36}^{+0.33}$&  $1.29_{-1.93}^{+2.31}$&   $-2.25_{-10.8}^{+1.35}$\\
    0.7 & $101.17_{-4.51}^{+5.28}$&    $-0.02_{-0.42}^{+0.46}$&  $1.21_{-1.53}^{+2.66}$&   $-2.53_{-15.0}^{+0.10}$\\
    0.8 & $107.20_{-6.30}^{+8.10}$&     $0.06_{-0.45}^{+0.55}$&  $1.13_{-1.15}^{+2.85}$&   $-2.75_{-17.5}^{+0.75}$\\
    0.9 & $115.04_{-10.58}^{+10.81}$&   $0.18_{-0.55}^{+0.57}$&  $1.33_{-1.16}^{+2.88}$&   $-3.76_{-19.8}^{+1.61}$\\
    1.0 & $122.08_{-13.76}^{+15.36}$&   $0.21_{-0.55}^{+0.64}$&  $1.28_{-0.94}^{+3.29}$&   $-3.90_{-23.5}^{+1.70}$\\
    1.1 & $129.90_{-17.10}^{+20.40}$&   $0.27_{-0.55}^{+0.62}$&  $1.37_{-0.87}^{+3.08}$&   $-4.50_{-22.3}^{+2.25}$\\
    1.2 & $136.60_{-20.09}^{+25.87}$&   $0.29_{-0.54}^{+0.73}$&  $1.32_{-0.73}^{+3.59}$&   $-4.50_{-26.9}^{+2.17}$\\
    1.3 & $146.50_{-23.85}^{+25.97}$&   $0.36_{-0.54}^{+0.63}$&  $1.45_{-0.78}^{+3.03}$&   $-5.30_{-22.1}^{+3.03}$\\
    1.4 & $154.90_{-25.35}^{+34.45}$&   $0.38_{-0.50}^{+0.71}$&  $1.44_{-0.71}^{+3.57}$&   $-5.40_{-27.0}^{+3.07}$\\
 \hline
 \hline
 \end{tabular}
 \caption{\label{parameters_SN} Estimates of $H,\,q,\,j,\,{\rm and}\,s$ as function of the redshift obtained from Union 2.1 SNe Ia sample for the 
 4th order approximation of $d_L$ (\ref{Dl-a-general}). The errors correspond to a 
 $2\sigma\,(\Delta\chi^2=4)$ statistical uncertainty for each parameter.  
 $\chi^2_{\nu}=\chi^2_{min}/{\rm NDoF}=0.976$ for all redshifts considered.}
 \end{table}

When we stop the $d_L$ expansion in the third term (Figs. \ref{3rdH_and_q} and \ref{3rdj}), a general feature is that SNe Ia constraints are tighter than the constraints obtained from $H(z)$ measurements in all redshift range covered. Particularly, the constraints on $j$ obtained from SNe Ia data are significantly stronger than the constraints obtained from $H(z)$ data. The constraints on $H$ and $q$ obtained from SNe Ia and from $H(z)$ data are compatible with the $\Lambda$CDM model and compatible with each other. For SNe Ia, values of $q>0$ are allowed for $z\geq0.5$ and values of $q<0$ are allowed for $z\leq0.8$, 
indicating that the transition between the decelerated to accelerated phases should stay in the range $0.5<z_t<0.8$. In turn, for $H$ data, positive values of $q$ are allowed for $z\geq0.3$ showing that in this case the 
transition redshift, $z_t$, is greater than $0.3$. The constraints on $j$ obtained from both, SNe Ia and $H(z)$ data are incompatible with the $\Lambda$CDM model. For SNe Ia data, $j$ begin to depart from the $\Lambda$CDM model at $z>0.3$. For $H(z)$ data $j$ is above the $\Lambda$CDM value, $j=1$, at $z=0.3$ and $z=0.4$ and below this value for $z\geq 1.2$. The constraints on $j$ reveals yet that the results obtained from $H(z)$ and SNe Ia data are incompatible with each other. 

For the fourth order expansion of $d_L$, the constraints from SNe Ia data, as expected, becomes weaker (Figs. \ref{H_and_q} and \ref{j_and_s}). For $z<1$ the constraints on $H$ obtained from SNe Ia data are tighter than the constraints obtained from $H$ measurements, reversing the roles for $z\geq 1$. A similar behavior is observed for $q$, with SNe Ia providing tighter constraints for $z\leq0.5$. For $j$ the constraints obtained from $H(z)$ data are tighter than the constraints provided by the SNe Ia data for $z\ge1$, while the constraints on $s$ obtained from $H$ data are tighter than those obtained from SNe Ia data for $z\geq0.8$. The constraints on $j$ obtained from $H$ data begin to depart from those from SN e Ia data for $z>0.8$, going to negative values. For the snap, the difference between the results obtained from SNe Ia and $H(z)$ data begins at $z>0.5$. 

As we can see, for this case, the results obtained from SNe Ia data are in agreement with the $\Lambda$CDM 
bounds, but still remains incompatible with the results obtained from $H$ data. Therefore, the inclusion of the fourth order term in the expansion of $d_L$ does not alleviate the tension between the data sets observed early. These results indicate a discrepancy between the 
 $H$ and SNe Ia data sets. Such a discrepancy cannot be seen when we restrict our analysis to the neighborhood of 
$z=0$. At $z=0$, the constraints on the parameters $j$ and $s$ are completely without statistical significance. 
Therefore, the standard cosmographic 
approach, which consists in expanding the Taylor series of $H$ and $d_L$ around $z=0$, does not seem a useful tool 
for testing models designed to explain the cosmic acceleration. This result is in agreement 
with the findings of \cite{delaCruz}. However, since their results 
remain valid regardless of the underlying cosmology, performing the series expansion around an arbitrary 
$\tilde{a}\neq1$ cosmography can still be an efficient way to rule out cosmological models. For instance,  a single
value of $j\neq1$ for some $z\neq0$ should be considered as evidence against the $\Lambda$CDM model.

It is important to note that, when we consider the fourth order expansion of $d_L$, at $z=0$, both, SNe Ia and $H(z)$ results do not exclude a decelerated Universe, $q_0>0$. 
However, it is an observational fact that, at the present time, the Universe is expanding at an accelerated rate 
\cite{Riess, Perlmutter}, i.e., $q_0<0$. So, how can we explain such a result? 
For SNe Ia data, this result can be explained by the fact that we are working with more terms 
in the $d_L$ expansion than necessary. When the expansion of $d_L$ is truncated at the most statistically 
significant term, we have $q_0<0$ at $2\sigma$ (see Table \ref{chi_square_cosmography}). Since, for $H$ data, 
we are already using the most relevant approximation, we suspect that this result may be due the low number 
of $H$ measurements or to the lack of precision of these measurements, or both.

Also note that, for the case of a fourth order expansion of $d_L$, values of 
$q<0$ are allowed in the entire redshift interval considered, i. e., both SNe Ia and $H$ data sets 
are compatible with an early time accelerated Universe. In this case, for SNe Ia, values of $q>0$ are allowed for $z\geq0.5$, 
indicating that the transition between the decelerated to accelerated phases should occur for redshifts greater 
than $0.5$. 

Also, we observe that from $z\geq0.6$ onwards the constraints on the snap obtained from SNe Ia data begin to 
become incompatible with the constraints coming from $H$ data. This confirms that we cannot combine the two data sets 
to reconstruct the time-dependence of the cosmographic parameters.

Finally, it should be mentioned that, even working with more parameters than necessary 
(which can be seem as a conservative analysis), the constraints obtained from SNe Ia data barely touch the $\Lambda$CDM 
diagnostic line $j=1$. That is, although compatible with the results, the $\Lambda$CDM is not the model most consistent with the data. 

\subsection{Transition redshift}

Although our results allow us to estimate the transition redshift $z_t$ by mere inspection of right panels of Figures \ref{3rdH_and_q} and \ref{H_and_q}, we want make it more precise. In ref. \cite{Moresco2016} it was noticed that the function $f(z)\equiv H(z)/(1+z)$ has an absolute minima at $z_t$. Then, building $f(z)$ from $H(z)$ data and fitting it with a piecewise linear function composed of two intervals (one for acceleration and one for deceleration), the authors were able to obtain a model-independent determination of $z_t$. By following this approach, we use the estimates of $H$ contained in the Tables \ref{parameters_H}, \ref{parameters_SN1} and \ref{parameters_SN} to estimate $z_t$. 

\nonumber For the sake of comparison, we also fit the open $\Lambda$CDM model

\begin{equation}
\label{olcdm}
H^2=H_0^2[\Omega_{m,0}(1+z)^3+\Omega_{k,0}(1+z)^2+\Omega_{\Lambda,0}],
\end{equation}
for which $z_t=[2\Omega_{\Lambda,0}/\Omega_{m,0}]^{1/3}-1$. 

Since the o$\Lambda$CDM model has two parameters less than the piecewise linear function, we use the corrected Akaike Information Criterion (AIC$_{{\rm C}}$) \cite{AICc}, and the Bayesian Information Criterion (BIC) \cite{BIC} to provide a fair comparison of the fits. These informations criteria are defined, respectively, as:

\begin{equation}
\label{AICc}
AIC_C\equiv-2\ln {\mathcal L}_{max} +\frac{2kN}{N-k-1}
\end{equation}
and
\begin{equation}
\label{BIC}
BIC\equiv-2\ln {\mathcal L}_{max} +k\ln N,
\end{equation}

\noindent where $k$ is the number of parameters of a given model and $N$ the Number of data point. 

\begin{table}[tpb]
\centering
\begin{tabular}{llccc}
\hline
\hline
Model&Set & $z_t$ & AIC$_{{\rm C}}$& BIC\\
\hline
o$\Lambda$CDM&$1$&$0.59^{+0.29}_{-0.09}$&5.46 &5.91\\
&$2$&$0.64^{+0.29}_{-0.10}$&12.52 &12.92\\
&$3$&$0.61^{+0.33}_{-0.13}$&9.8 &10.23\\
\hline
Pice-wise&$1$&$0.34^{+0.40}_{-0.29}$& 12.27&11.10\\
&$2$&$0.36^{+0.24}_{-0.14}$& 13.64&12.47\\
&$3$&$0.35^{+0.35}_{-0.22}$& 12.82&11.65\\
\hline
\hline
\end{tabular}
\caption{\label{transition_redshift_f} Constraints on the cosmological transition redshift$z_t$ obtained by fitting $f=H/(1+z)$ estimates obtained from Tables \ref{parameters_H} (set 1), \ref{parameters_SN1} (set 2) and \ref{parameters_SN} (set 3) with (\ref{g_exp}) and o$\Lambda$CDM model (\ref{olcdm}). The errors correspond to 95\% confidence level.}
\end{table}

Table \ref{transition_redshift_f} contain the constraints on $z_t$ at $2\sigma$ confidence level. The sets 1, 2, and 3 refers to estimates of $f(z)$ obtained from Tables \ref{parameters_H}, \ref{parameters_SN1} and \ref{parameters_SN}, respectively. Our results are compatible with the findings of \cite{Moresco2016} that constrain the transition redshift at $1\sigma$ confidence level to $0.3\leq z_t\leq0.5$ for a piecewise linear function fit and $0.58\leq z_t\leq74$ for the o$\Lambda$CDM model.
AIC$_{{\rm C}}$ and BIC estimators reveal that $H(z)$ data (set 1) provides strong evidence in favor of the o$\Lambda$CDM model ($\Delta AIC_C, \Delta BIC>5$) while SNe Ia data (sets 2 and 3) do not favor any of the models considered\footnote{in fact, the set 3 provides $\Delta AIC_C>2.5$, which is a significant evidence in favor of o$\Lambda$CDM model, but $\Delta BIC<2.5$ which is a weak evidence}. 


Now, instead of use $f(z)$, we can constrain $z_t$ with our estimates of $q$ by building the function $g(z)\equiv f' /f= q(z)/(1+z)$. Since $g(z_t)=0$, it is natural try to adjust $g$ by a second order expansion, i. e.,

\begin{equation}
\label{g_exp}
g(z)=g'(z_t)(z-zt)+\frac{1}{2}g''(z_t)(z-z_t)^2.
\end{equation}

\begin{figure}[tbp]
\centering
\includegraphics[width=15cm,height=8cm]{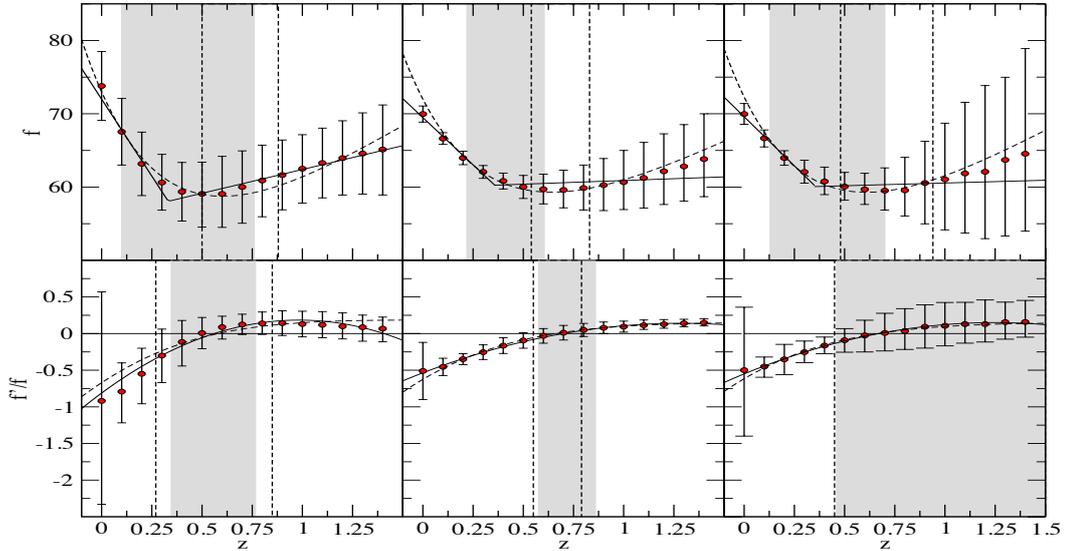}
\caption{\label{z_t} The functions $f=H/(1+z)$ (top panels) and $g=q/(1+z)$ (bottom panels) obtained from Tables \ref{parameters_H} (left), \ref{parameters_SN1} (center) and \ref{parameters_SN} (right). The solid curves corresponds to the best fit of a piecewise linear function (top panels) and the function $g(z)$ given by (\ref{g_exp}) (bottom panels). The dashed curves corresponds to the o$\Lambda$CDM model. The vertical grey regions denotes the constraint on the transition redshift for the piecewise linear function (top panels) and the fitting function (\ref{g_exp}) (bottom panels). The vertical dashed lines denotes the constraint on $z_t$ for the o$\Lambda$CDM model. The horizontal solid line in the bottom panels marks the transition from the decelerated to the accelerated phase. Since $g(z_t)=0$, we can estimate the transition from decelerated to accelerated expansion phase by eye in the bottom panels.}
\end{figure}

\noindent Since our estimates of $g(z)$ are cosmology-independent, we should presume that the estimate of $z_t$ achieved in this way it is also cosmology-independent. Table \ref{transition_redshift} contain the constraints on $z_t$ at $2\sigma$ confidence level for this case.
AIC$_{{\rm C}}$ and BIC estimators reveal that SNe Ia data (sets 2 and 3) favor the o$\Lambda$CDM model ($\Delta AIC_C, \Delta BIC>2.5$) while $H(z)$ data (set 1) do not favor any of the fitting functions considered. These results confirms what we have already noticed. Note that the weak constraints on $z_t$ from set 3 can be due the unnecessary term include in the $d_L$ approximation. 
Figure \ref{z_t} shows the functions $f$ (top panels) and $g=f'/f$ (bottom panels) obtained from Tables \ref{parameters_H} (left), \ref{parameters_SN1} (center) and \ref{parameters_SN} (right). The solid curve corresponds to the best fit of piecewise linear function (top panels) and $g(z)$ function given by (\ref{g_exp}). The dashed curve corresponds to the o$\Lambda$CDM model. The vertical grey region is the constraint on $z_t$ for the piecewise linear function and for the polynomial fit (\ref{g_exp}). The vertical dashed lines denotes the constraint on $z_t$ for the o$\Lambda$CDM model. The horizontal solid line in the bottom panels marks the transition from the decelerated to the accelerated phase.  By following the vertical stripes we can see that the constraints on $z_t$ for the o$\Lambda$CDM model from both $f$ and $g$ estimates are entirely compatible with each other. Also, the o$\Lambda$CDM constraints are entirely compatible with the model-independent constraints on $z_t$ provide for the polynomial fit (\ref{g_exp}), but are not in good agree with the piecewise bounds on $z_t$.

\begin{table}[tpb]
\centering
\begin{tabular}{llccc}
\hline
\hline
Model&Set & $z_t$ & AIC$_{{\rm C}}$& BIC\\
\hline
o$\Lambda$CDM&$1$&$0.58^{+0.27}_{-0.31}$&9.73 &10.15\\
&$2$&$0.66^{+0.13}_{-0.11}$&5.32 &5.74\\
&$3$&$0.67^{}_{-0.22}$&5.16 &5.58\\
\hline
Polynomial&$1$&$0.56^{+0.21}_{-0.21}$& 10.01&9.95\\
&$2$&$0.70^{+0.16}_{-0.12}$&8.78 &8.73\\
&$3$&$0.67^{}_{-0.21}$&8.28 &8.22\\
\hline
\hline
\end{tabular}
\caption{\label{transition_redshift} Constraints on the cosmological transition redshift$z_t$ obtained by fitting $g=q/(1+z)$ estimates obtained from Tables \ref{parameters_H} (set 1), \ref{parameters_SN1} (set 2) and \ref{parameters_SN} (set 3) with (\ref{g_exp}) and o$\Lambda$CDM model (\ref{olcdm}). The errors correspond to 95\% confidence level.}
\end{table}

\section{Final remarks}

In this paper we have used the cosmographic approach to constrain the Hubble ($H$), deceleration ($q$), jerk ($j$) and snap ($s$) parameters at $z\ne0$ from SNe Ia and Hubble parameter data. These constraints are obtained from data by changing the expansion center of the $H$ and $d_L$ Taylor series at small intervals. Such simple implementation allows us to map the time evolution of the cosmographic parameters without assuming a specific gravity model or making assumptions about the sources. This approach can be a useful tool to decide between modified gravity or dark energy models designed to explain the current accelerated expansion of the Universe. For instance, for the main candidate used to explain the present cosmic acceleration, the $\Lambda$CDM model, $j=1$. In the usual approach, where the expansion center is fixed at $z=0$, evidence against the $\Lambda$CDM model is possible only if we find $j_0\neq1$ with some statistical significance. However, many cosmographic analyses performed with multiple data sets have shown that the constraints on $j_0$ are too weak and do not allow us to decide either for or against $\Lambda$CDM (or many other competing models). On the other hand, in the method used in this paper, it is enough to find a single value of $j\neq1$ with some statistical significance to rule out the $\Lambda$CDM model.

For both, SNe Ia and $H(z)$ data, we show that the value $j=1$ is rule out at $2\sigma$ confidence level when we stop the series of $d_L$ and $H$ at the last relevant term. This result put difficulties on the $\Lambda$CDM model. Our results also indicates that SNe Ia and $H(z)$ data are incompatible with each other. When we take a fourth order expansion for $d_L$ expansion, the SNe Ia data accommodate the $\Lambda$CDM model. In this case,  the constraints on the cosmographic parameters obtained from SNe Ia data are weaker than they should be. Even so, the $2\sigma$ bounds do not overlap and the results obtained from SNe Ia data remains incompatible with results obtained from $H(z)$ data. 

These conflicting results may indicates a tension between SNe Ia and $H(z)$ data, which is masked at $z=0$. Such a discrepancy indicates that we cannot combine these two data sets to reconstruct the time evolution of the kinematic parameters. In fact, the Taylor series of $H$ and $d_L$ cannot be treated on equal footing since we need to include more terms than necessary in the $d_L$ approximation to make a combination possible. If we look at the results of SNe Ia and $H(z)$ data separately, we will conclude that the $\Lambda$CDM model is excluded. However we cannot make such an extreme statement since both, the results of SNe Ia and $H(z)$ data, are not in agreement with each other. 
We believe that future analyses with a larger and more accurate $H$ data can help us to clarify this problem.

\begin{acknowledgments}

CRF acknowledge the financial support from Coordena\c{c}\~ao de Aperfei\c{c}oamento de Pessoal de N\'{i}vel Superior (CAPES). The authors acknowledge Thomas Dumelow and Jailson Alcaniz for useful comments.

\end{acknowledgments}


\begin{thebibliography}{99}

\bibitem{Riess} A. Riess et al., {\it Observational Evidence from Supernovae for an Accelerating Universe and a Cosmological Constant, Astron.J.} {\bf 116} (1998) 1009

\bibitem{Perlmutter} S. Perlmutter et al., {\it Measurements of Ω and Λ from 42 High-Redshift Supernovae, Astrophys.J.} {\bf 517} (1999) 565

\bibitem{LCDM} T. Padmanabhan, {\it Cosmological Constant - the Weight of the Vacuum, Phys. Rept.} {\bf 380} (2003) 235

\bibitem{CC_problem} S. Weinberg, {\it The Cosmological Constant Problem, Rev. Mod. Phys.} {\bf 61} (1989) 1

\bibitem{fr1} S. Nojiri and S. D. Odintsov, {\it Where new gravitational physics comes from: M Theory?, Phys. Lett. B} {\bf 576} (2003) 5

\bibitem{fr2} L. Amendola, D. Polarski D. and S. Tsujikawa, {\it Are f(R) dark energy models cosmologically viable ?, Phys. Rev. Lett.} {\bf 98} (2007) 131302

\bibitem{fr3} S. Capozziello and A. Felice, {\it f(R) cosmology by Noether's symmetry, JCAP} {\bf0808} (2008) 016

\bibitem{brane1} L. Randall and R. Sundrum, {\it An Alternative to compactification, Phys. Rev. Lett.} {\bf83} (1999) 3370

\bibitem{brane2} G. Dvali, G. Gabadadze and M. Porrati, {\it 4-D gravity on a brane in 5-D Minkowski space, Phys. Lett. B} {\bf485} (2000) 208

\bibitem{brane3} C. Deffayet, {\it Cosmology on a brane in Minkowski bulk, Phys. Lett. B} {\bf 502} (2001) 199

\bibitem{brane4} R. Dick, {\it Brane worlds, Class. Quant. Grav.} {\bf 18} (2001) R1

\bibitem{quint} C. Wetterich, {\it Cosmology and the fate of dilatation symmetry, Nucl. Phys. B} {\bf 302} (1988) 668

\bibitem{phant} R. R. Caldwell, {\it A Phantom menace?, Phys. Lett. B} {\bf 545} (2002) 23

\bibitem{quintom} M. R. Setare and E. N. Saridakis, {\it Quintom model with O(N) symmetry, JCAP} {\bf0809} (2008) 026

\bibitem{zcosmog1} T. Chiba and T. Nakamura, {\it The Luminosity distance, the equation of state, and the geometry of the universe, Prog. Theor. Phys.} {\bf 100} (1998) 1077

\bibitem{zcosmog2} M. Visser, {\it Jerk and the cosmological equation of state, Class. Quant. Grav.}  {\bf 21} (2004) 2603.

\bibitem{zcosmog3} M. Visser, {\it Cosmography: Cosmology without the Einstein equations, Gen. Rel. Grav.}  {\bf 37} (2005) 1541.

\bibitem{cosmography1} D. Rapetti,  S. W. Allen, M. A. Amin and R. D. Blandford, {\it A kinematical approach to dark energy studies, MNRAS} {\bf 375} (2007) 1510

\bibitem{cosmography2} M. S. Turner and A. G. Riess, {\it Do SNe Ia provide direct evidence for past deceleration of the universe?, Astrophys.J.} {\bf 569} (2002) 18

\bibitem{cosmography3} L. Alam,  V. Sahni, T. Deep Saini and A. A. Starobinsky, {\it Exploring the expanding universe and dark energy using the Statefinder diagnostic, MNRAS} {\bf 344} (2003) 1057

\bibitem{Cattoen2007a} C. Catt\"oen and M. Visser, {\it The Hubble series: Convergence properties and redshift variables, Class. Quant. Grav.} 24 (2007) 5985

\bibitem{Cattoen2007b} C. Catt\"oen and Visser, {\it Cosmography: Extracting the Hubble series from the supernova data}, gr-qc/0703122

\bibitem{Barboza2012} E. M. Barboza Jr. and F. C. Carvalho, {\it A kinematic method to probe cosmic acceleration, Phys. Lett. B} 715 (2012) 19

\bibitem{Vitagliano} V. Vitagliano, J.-Q. Xia, S. Liberati and M. Viel, {\it High-Redshift Cosmography, JCAP} {\bf 1003} (2010) 005

\bibitem{Planck} P. A. R. Ade et al.: Planck Collaboration, {\it Planck 2015 results. XIII. Cosmological parameters, Astron. Astrophys.} {\bf 594} (2015) A13 

\bibitem{Union2.1} N. Suzuki et al.: The supernova Cosmology Project, {\it The Hubble Space Telescope Cluster Supernova Survey: V. Improving the Dark Energy Constraints Above $z>1$ and Building an Early-Type-Hosted Supernova Sample, Astrophys. J.}  {\bf 746} (2012) 85.

\bibitem{Moresco2016} M. Moresco, et al.,{\it A 6\% measurement of the Hubble parameter at $z\sim0.45$:  direct evidence of the epoch of cosmic re-acceleration, JCAP} {\bf 1605} (2016) 014

\bibitem{RiessH_0} A. Riess et al., {\it A 2.4\% Determination of the Local Value of the Hubble Constant,  Atrophys. J.} {\bf 826} (2016) 56

\bibitem{delaCruz} V. C. Busti,  A. Cruz-Dombriz, P. K. S. Dunsby and D. S\'aez-G\'omez, {\it Is cosmography a useful tool for testing cosmology?, Phys. Rev D} {\bf 92} (2015) 123512

\bibitem{AICc} N. Sugiura, {\it Further analysis of the data by Akaikes information criterion and the finite corrections,
Communications in Statistics - Theory and Methods} {\bf A7} (1978) 13.

\bibitem{BIC} G. Schwarz, {\it Estimating the dimension of a model, Ann. Statist.} {\bf6} (1978) 461.


%
%
%
%
%
%
         

%
%
%
%
%
%


\end{thebibliography}
\end{document}